\documentclass[twocolumn,english]{revtex4}
\usepackage[T1]{fontenc}
\usepackage[latin9]{inputenc}
\usepackage{color}
\usepackage{array}
\usepackage{multirow}
\usepackage{amsmath}
\usepackage{amssymb}
\usepackage{graphicx}

\makeatletter

\newcommand*\LyXZeroWidthSpace{\hspace{0pt}}
\providecommand{\tabularnewline}{\\}

\@ifundefined{textcolor}{}
{%
 \definecolor{BLACK}{gray}{0}
 \definecolor{WHITE}{gray}{1}
 \definecolor{RED}{rgb}{1,0,0}
 \definecolor{GREEN}{rgb}{0,1,0}
 \definecolor{BLUE}{rgb}{0,0,1}
 \definecolor{CYAN}{cmyk}{1,0,0,0}
 \definecolor{MAGENTA}{cmyk}{0,1,0,0}
 \definecolor{YELLOW}{cmyk}{0,0,1,0}
}

\makeatother

\usepackage{babel}
\begin{document}
\title{Natural orbital functional for multiplets}
\author{Mario Piris}
\address{Donostia International Physics Center (DIPC), 20018 Donostia, Euskadi,
Spain;}
\address{IKERBASQUE, Basque Foundation for Science, 48013 Bilbao, Euskadi,
Spain.\bigskip{}
}
\begin{abstract}
A natural orbital functional for electronic systems with any value
of the spin is proposed. This energy functional is based on a new
reconstruction for the two-particle reduced density matrix (2RDM)
of the multiplet, that is, of the mixed quantum state that allows
all possible spin projections. The mixed states of maximum spin multiplicity
are considered. This approach differs from the methods routinely used
in electronic structure calculations that focus on the high-spin component
or break the spin symmetry. In the ensemble, there are no interactions
between electrons with opposite spins in singly occupied orbitals,
as well as new inter-pair $\alpha\beta$-contributions are proposed
in the 2RDM. The proposed 2RDM fulfills (2,2)-positivity necessary
N-representability conditions and guarantees the conservation of the
total spin. The NOF for multiplets is able to recover the non-dynamic
and the intrapair dynamic electron correlation. The missing dynamic
correlation is recovered by the NOF-MP2 method. Calculation of ionization
potentials of the first-row transition-metal atoms is presented as
test case. The values \LyXZeroWidthSpace \LyXZeroWidthSpace obtained
agree with those reported at the coupled cluster singles and doubles
level of theory with perturbative triples and experimental data.
\end{abstract}
\maketitle

\section{Introduction}

In the absence of an external field at absolute zero temperature,
the non-relativistic Hamiltonian used in electronic calculations does
not contain spin coordinates. The eigenvectors of such a Hamiltonian
are eigenvectors of the square of the total spin ($\hat{S}^{2}$)
and of one of its components, usually chosen to be $\hat{S}_{z}$.
Consequently, the ground state of a many-electron system with total
spin $S$ is a multiplet, i.e., a mixed quantum state that allows
all possible $S_{z}$ values.

Despite its success, the Kohn-Sham density functional theory \citep{Kohn1965},
including the spin-dependent-density formalism, can not describe the
degeneracy of the spin-multiplet components. The proper description
of the ensemble of pure spin states is provided by the one-particle
reduced density matrix (1RDM) functional theory \citep{Gilbert1975}.
In 1980 \citep{Valone1980}, Valone extended Levy's functional \citep{Levy1979}
to include all ensemble N-representable 1RDMs in its domain. A series
of conditions fulfilled by this exact functional have been recently
investigated \citep{Rohr2011}. Unfortunately, computational schemes
based on the constrained search formulation are several times more
expensive than solving directly the Schrödinger equation; therefore,
the construction of the functional requires a practical approach.

Actually, we must only construct the 1RDM functional for the electron-electron
potential energy $V_{ee}$, since the non-interacting part of the
electronic Hamiltonian is a one-particle operator. $V_{ee}$ is a
well-known functional of the two-particle reduced density matrix (2RDM),
hence a convenient approach is to reconstruct the 2RDM in terms of
the 1RDM. However, this does not fully reconstruct the ground-state
energy, only the 2RDM. Approximating the energy implies that theorems
established for the exact 1RDM functional may not be fulfilled since
the 2RDM dependence remains \citep{Donnelly1979}. To guarantee the
existence of an N-electron system compatible with a proposed functional,
we must observe the N-representability conditions \citep{Mazziotti2012a}
on the reconstructed 2RDM. Therefore, we are dealing with approximate
1RDM methods, where the 2RDM continues to play a dominant, albeit
hidden, role \citep{Piris2018}.

In most applications, the spectral decomposition of the 1RDM is used
to express it in terms of the naturals orbitals (NOs) and their occupation
numbers (ONs). In this representation, the energy expression is referred
to as natural orbital functional (NOF). Within NOF approximations,
spin-dependent formalisms have been also proposed \citep{Goedecker1998,Lathiotakis2005},
which do not conserve spin. NOFs that correctly reproduce the expectation
values of spin operators have been reported \citep{Leiva2007,Piris2009,Piris2010,Quintero-Monsebaiz2019},
however, only for the high-spin component of the multiplet.

The aim of this work is to propose a NOF that allows to describe an
electronic system with any value of $S$, that is, a new reconstruction
of the 2RDM for spin-multiplets. For the latter, the mixed states
of maximum multiplicity are considered.

The chemical versatility of transition metal (TM) atoms is derived
from their ability to exist in multiple spin states, so there is considerable
interest in having the ability to predict reliable properties for
these systems. One of these properties is the ionization potential
(IP). The correct description of TM IP is a great challenge since
not only a significant rearrangement of the electronic configurations
takes place, but the resulting contraction of the electron cloud implies
a considerable contribution of the dynamic correlation. Hence, multi-reference
methods are needed for an accurate description \citep{Thomas2015}.
Unfortunately, multireference methods are more complicated in implementation
and much less easy to use than single reference methods.

Recently \citep{Piris2017,Piris2018b}, a single-reference method
for the electron correlation was introduced taking as reference the
Slater determinant formed with the NOs of an approximate NOF. In this
approach, called \textcolor{black}{natural orbital functional - second-order
Møller\textendash Plesset (NOF-MP2)} method, the total energy of an
N-electron system can be attained by the expression
\begin{equation}
E=\tilde{E}_{hf}+E^{corr}=\tilde{E}_{hf}+E^{dyn}+E^{sta}\label{Etotal}
\end{equation}

where $\tilde{E}_{hf}$ is the Hartree-Fock (HF) energy obtained with
the NOs, the dynamic energy ($E^{dyn}$) is derived from a modified
MP2 perturbation theory, while the non-dynamic energy ($E^{sta}$)
is obtained from the static component of the employed NOF. An important
feature of the method is that double counting is avoided by taking
the amount of static and dynamic correlation in each orbital as a
function of its occupancy.

The second aim of the present work is to extend the NOF-MP2 method
to the spin multiplets and apply it to the description of the first-row
TM IPs.

This article is organized as follows. We start in section II with
the basic concepts and notations related to the NOF theory. Section
III is devoted to the extension of the NOF-MP2 method to spin multiplets.
Finally, I present the results obtained for the IPs of the TM Sc-Zn
elements in Section IV.

\section{Natural Orbital Functional Theory}

Consider the N-electron Hamiltonian
\begin{equation}
\hat{H}=\sum\limits _{ik}\mathcal{H}_{ki}\hat{a}_{k}^{\dagger}\hat{a}_{i}+\frac{1}{2}\sum\limits _{ijkl}\left\langle kl|ij\right\rangle \hat{a}_{k}^{\dagger}\hat{a}_{l}^{\dagger}\hat{a}_{j}\hat{a}_{i}\label{Ham}
\end{equation}
where $\mathcal{H}_{ki}$ denote the matrix elements of the one-particle
part of the Hamiltonian involving the kinetic energy and the potential
energy operators, $\left\langle kl|ij\right\rangle $ are the two-particle
interaction matrix elements, $\hat{a}_{i}^{\dagger}$ and $\hat{a}_{i}$
are the familiar fermion creation and annihilation operators associated
with the complete orthonormal spin-orbital set $\left\{ \left|i\right\rangle \right\} $.
For a given spin $S$, there are $\left(2S+1\right)$ energy degenerate
eigenvectors $\left|SM\right\rangle $, so a mixed state is defined
by the N-particle density matrix statistical operator
\begin{equation}
\mathfrak{\hat{D}}={\displaystyle {\displaystyle {\textstyle {\displaystyle \sum_{M=-S}^{S}}}}\omega_{M}\left|SM\right\rangle \left\langle SM\right|}\label{DM}
\end{equation}
In Eq. (\ref{DM}), $\omega_{M}$ are positive real numbers that sum
one, so that $\mathfrak{\hat{D}}$ corresponds to a weighted sum of
all accessible pure states. For equiprobable pure states, we take
$\omega_{M}=(2S+1)^{-1}$.

The expectation value of (\ref{Ham}) reads as
\begin{equation}
E=\sum\limits _{ik}\mathcal{H}_{ki}\Gamma_{ki}+\sum\limits _{ijkl}\left\langle kl|ij\right\rangle D_{kl,ij}\label{Energy}
\end{equation}
where the 1RDM and 2RDM are
\begin{equation}
\begin{array}{c}
\Gamma_{ki}={\displaystyle {\textstyle {\displaystyle \sum_{M=-S}^{S}}}}\omega_{M}\left\langle SM\right|\hat{a}_{k}^{\dagger}\hat{a}_{i}\left|SM\right\rangle \\
D_{kl,ij}={\displaystyle {\textstyle {\displaystyle \frac{1}{2}\sum_{M=-S}^{S}}}}\omega_{M}\left\langle SM\right|\hat{a}_{k}^{\dagger}\hat{a}_{l}^{\dagger}\hat{a}_{j}\hat{a}_{i}\left|SM\right\rangle 
\end{array}
\end{equation}
Normalization of Löwdin is used, in which the traces of the 1RDM and
2RDM are equal to the number of electrons and the number of electron
pairs, respectively. The last term in Eq. (\ref{Energy}) is $V_{ee}$,
an explicit functional of the 2RDM. To construct the functional $V_{ee}\left[\Gamma\right]$,
we employ the representation where the 1RDM is diagonal ($\hat{\Gamma}={\displaystyle {\textstyle \sum}_{i}n_{i}}\left|i\right\rangle \left\langle i\right|$).
Restriction on the ONs to the range $0\leq n_{i}\leq1$ represents
a necessary and sufficient condition for ensemble N-representability
of the 1RDM \citet{Coleman1963}. This leads to a NOF, namely:
\begin{equation}
E=\sum\limits _{i}n_{i}\mathcal{H}_{ii}+\sum\limits _{ijkl}D[n_{i},n_{j},n_{k},n_{l}]\left\langle kl|ij\right\rangle \label{ENOF}
\end{equation}
where $D[n_{i},n_{j},n_{k},n_{l}]$ represents the reconstructed ensemble
2RDM from the ONs. For $\hat{S}_{z}$ eigenvectors, density matrix
blocks that conserve the number of each spin type are non-vanishing,
however, only three of them are independent, namely $D^{\alpha\alpha}$,
$D^{\alpha\beta}$, and $D^{\beta\beta}$ \citep{Piris2007b}.

Consider $\mathrm{N_{I}}$ single electrons and $\mathrm{N_{II}}$
paired electrons, so that $\mathrm{N_{I}}+\mathrm{N_{II}}=\mathrm{N}$.
Assume further that all spins corresponding to $\mathrm{N_{II}}$
electrons are coupled as a singlet, thence the $\mathrm{N_{I}}$ electrons
determine the spin of the system. In fact, excluding the singlet part,
the $\hat{S}_{z}$ eigenfunctions $\theta_{k}$ are simply products
of $\mathrm{N_{I}}$ one-electron functions ($\alpha$'s or $\beta$'s):
\begin{equation}
\theta_{k}=\sigma\left(1\right)\sigma\left(2\right)...\sigma\left(\mathrm{N_{I}}\right),\quad\sigma\left(i\right)=\left\{ \begin{array}{c}
\alpha\left(i\right)\\
\beta\left(i\right)
\end{array}\right.\label{primi}
\end{equation}
\begin{equation}
\hat{S}_{z}\theta_{k}=\frac{1}{2}\left(\mu-\nu\right)\theta_{k}
\end{equation}

where $\mu$ is the number of $\alpha$'s, and $\nu$ is the number
of $\beta$'s, respectively. For a given $\mu$, we have $\left(\begin{array}{c}
\mathrm{N_{I}}\\
\mu
\end{array}\right)=\left(\begin{array}{c}
\mathrm{N_{I}}\\
\nu
\end{array}\right)$ primitive spin functions (\ref{primi}) which belong to the same
eigenvalue of $\hat{S}_{z}$. For instance, we have four primitive
spin functions in the two-electron case, namely,
\begin{equation}
\alpha\left(1\right)\alpha\left(2\right),\alpha\left(1\right)\beta\left(2\right),\beta\left(1\right)\alpha\left(2\right),\beta\left(1\right)\beta\left(2\right)
\end{equation}
Let us define the square brackets of spins functions, $\left[\alpha^{\mu}\beta^{\nu}\right]$,
as the sum of all primitive spin functions with $\mu$ $\alpha$-spin
functions and $\nu$ $\beta$-spin functions. Example:
\begin{equation}
\left[\alpha^{2}\beta^{2}\right]=\alpha\alpha\beta\beta+\alpha\beta\alpha\beta+\alpha\beta\beta\alpha+\beta\alpha\alpha\beta+\beta\alpha\beta\alpha+\beta\beta\alpha\alpha
\end{equation}
It is not difficult to demonstrate \citep{Paunez2000} that the square
brackets are simultaneous eigenfunctions of $\hat{S}^{2}$ and $\hat{S}_{z}$
operators:
\begin{equation}
\hat{S}^{2}\left[\alpha^{\mu}\beta^{\nu}\right]=\frac{\mathrm{N_{I}}}{2}\left(\frac{\mathrm{N_{I}}}{2}+1\right)\left[\alpha^{\mu}\beta^{\nu}\right]
\end{equation}
\begin{equation}
\hat{S}_{z}\left[\alpha^{\mu}\beta^{\nu}\right]=\frac{1}{2}\left(\mu-\nu\right)\left[\alpha^{\mu}\beta^{\nu}\right]
\end{equation}
Note that the eigenvalue $M=\left(\mu-\nu\right)/2$ can take values
\LyXZeroWidthSpace \LyXZeroWidthSpace in the range $-\mathrm{N_{I}}/2\leq M\leq\mathrm{N_{I}}/2$.
All the square brackets are simultaneous eigenfuntions of $\hat{S}^{2}$
and $\hat{S}_{z}$ with the quantum numbers $\mathrm{N_{I}}/2$ and
$M$, respectively. This is the mixed state of highest multiplicity:
$2S+1=\mathrm{N_{I}}+1$, and it is the only state that belongs to
the quantum number $S=\mathrm{N_{I}}/2$ \citep{Paunez2000}. For
example, for $\mathrm{N_{I}}=2$ the triplet state ($S=1$) is
\begin{equation}
\left\{ \begin{array}{c}
\qquad\qquad\alpha\left(1\right)\alpha\left(2\right),\qquad\qquad M=1\\
\frac{1}{\sqrt{2}}\left[\alpha\left(1\right)\beta\left(2\right)+\beta\left(1\right)\alpha\left(2\right)\right],\:M=0\\
\qquad\qquad\beta\left(1\right)\beta\left(2\right),\qquad\qquad M=-1
\end{array}\right.
\end{equation}

Interestingly, the expected value of $\hat{S}_{z}$ for the whole
ensemble is zero,
\begin{equation}
\mathrm{<}\hat{S}_{z}\mathrm{>}=\frac{1}{\mathrm{N_{I}}+1}{\textstyle {\displaystyle \sum_{M=-\mathrm{N_{I}}/2}^{\mathrm{N_{I}}/2}}M}=0
\end{equation}
In the absence of single electrons ($\mathrm{N_{I}}=0$), the energy
must obviously be reduced to a NOF that well describes singlet states.
Let us employ the electron-pairing approach in NOF theory \citep{Piris2018a}
for this purpose. First, the spin-restricted theory is adopted, in
which a single set of spatial orbitals $\left\{ \left|p\right\rangle \right\} $
is used for $\alpha$ and $\beta$ spins. Accordingly, all spatial
orbitals are double occupied in the ensemble, so that occupancies
for particles with $\alpha$ and $\beta$ spins are equal: $n_{p}^{\alpha}=n_{p}^{\beta}=n_{p}.$

Let us divide in turn the orbital space $\Omega$ into two subspaces:
$\Omega=\Omega_{\mathrm{I}}\oplus\Omega_{\mathrm{II}}$. The orbital
space $\Omega_{\mathrm{II}}$ is composed of $\mathrm{N_{II}}/2$
mutually disjoint subspaces: $\Omega_{\mathrm{II}}=\Omega_{1}\oplus\Omega_{2}\oplus...\oplus\Omega_{\mathrm{N_{II}}/2}$,
$\Omega_{f}\cap\Omega_{g}=\textrm{Ø}$ \citep{Piris2018a}. Each subspace
$\Omega{}_{g}\in\Omega_{\mathrm{II}}$ contains one orbital $g$ below
the level $\mathrm{N_{II}}/2$, and $\mathrm{N}_{g}$ orbitals above
it, so $dim\left\{ \Omega_{g}\right\} =\mathrm{N}_{g}+1$. In this
work, $\mathrm{N}_{g}$ is equal to a fixed number corresponding to
the maximum value allowed by the basis set used in calculations.

Taking into account the spin, the total occupancy for a given subspace
$\Omega{}_{g}$ is 2, which is reflected in additional sum rule, namely,
\begin{equation}
\sum_{p\in\Omega_{g}}n_{p}=1,\,\Omega{}_{g}\in\Omega{}_{\mathrm{II}}\label{sum1}
\end{equation}
It follows that
\begin{equation}
2\sum_{p\in\Omega_{\mathrm{II}}}n_{p}=2\sum_{g=1}^{\mathrm{N_{II}}/2}\sum_{p\in\Omega_{g}}n_{p}=\mathrm{N_{II}}\label{sumNp}
\end{equation}

Similarly, $\Omega_{\mathrm{I}}$ is composed of $\mathrm{N_{I}}$
mutually disjoint subspaces: $\Omega_{\mathrm{I}}=\Omega_{\mathrm{N_{II}}/2+1}\oplus...\oplus\Omega_{\mathrm{N_{\Omega}}}$.
Here, $\mathrm{\mathrm{N}_{\Omega}=}\mathrm{N_{II}}/2+\mathrm{N_{I}}$
denotes the total number of suspaces in $\Omega$. In contrast to
$\Omega_{\mathrm{II}}$, each subspace $\Omega{}_{g}\in\Omega_{\mathrm{I}}$
contains only one orbital $g$ with $\mathrm{N_{II}}/2<g\leq N_{\Omega}$,
i.e., $dim\left\{ \Omega_{g}\right\} =1$. Besides, we take $2n_{g}=1$,
so each orbital is fully occupied individually, but we do not know
whether the electron has $\alpha$ or $\beta$ spin: $n_{g}^{\alpha}=n_{g}^{\beta}=n_{g}=1/2$.
It follows that
\begin{equation}
2\sum_{p\in\Omega_{\mathrm{I}}}n_{p}=2\sum_{g=\mathrm{N_{II}}/2+1}^{\mathrm{N_{\Omega}}}n_{g}=\mathrm{N_{I}},
\end{equation}
so taking into account Eq. (\ref{sumNp}), the trace of the 1RDM is
verified equal to the number of electrons: 
\begin{equation}
2\sum_{p\in\Omega}n_{p}=2\sum_{p\in\Omega_{\mathrm{II}}}n_{p}+2\sum_{p\in\Omega_{\mathrm{I}}}n_{p}=\mathrm{N_{II}}+\mathrm{N_{I}}=\mathrm{\mathrm{N}}
\end{equation}
\textcolor{black}{}
\begin{figure}
\begin{centering}
\textcolor{black}{\caption{\label{fig1} Splitting of the orbital space $\Omega$ into subspaces.
In this example, $S=1$ (triplet) and $\mathrm{N_{I}}=2$, so two
orbitals make up the subspace $\Omega_{\mathrm{I}}$, whereas six
electrons ($\mathrm{N_{II}}=6$) distributed in three subspaces $\left\{ \Omega_{1},\Omega_{2},\Omega_{3}\right\} $
make up the subspace $\Omega_{\mathrm{II}}$. Note that the maximum
value allowed by the basis set is $\mathrm{N}_{g}=4$. The arrows
depict the values \LyXZeroWidthSpace \LyXZeroWidthSpace of the ensemble
occupation numbers, alpha ($\downarrow$) or beta ($\uparrow$), in
each orbital.\bigskip{}
}
}
\par\end{centering}
\centering{}\textcolor{black}{\includegraphics[scale=0.45]{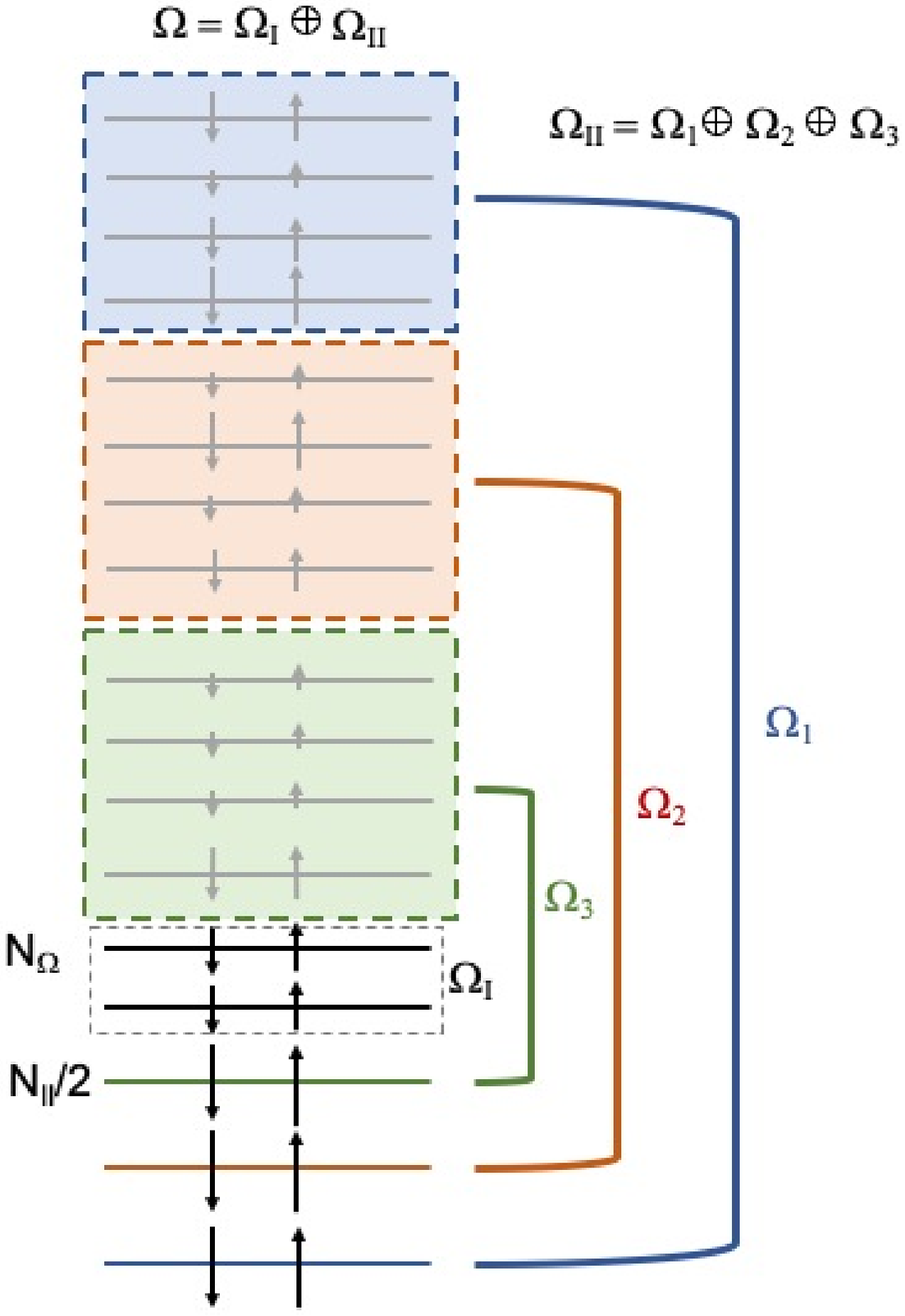}}
\end{figure}

It is worth noting that the splitting of the orbital space into the
subspaces $\Omega_{\mathrm{I}}$ and $\Omega_{\mathrm{II}}$ is determined
by the value of the total spin $S$. The latter determines the number
of orbitals in the subspace $\Omega_{\mathrm{I}}$, which is equal
to the number of unpaired electrons. If a system has $S=1/2$, only
one orbital composes the subspace $\Omega_{\mathrm{I}}$, analogously,
if a system has $S=1$, we have two orbitals in $\Omega_{\mathrm{I}}$,
for $S=3/2$, we have three orbitals in $\Omega_{\mathrm{I}}$, and
so on. The rest of the orbitals can be distributed in the subspace
$\Omega_{\mathrm{II}}$ that constitutes a singlet, and to which this
division is reduced in the case $S=0$ as expected.

The electron-pairing approach is used to divide the subspace $\Omega_{\mathrm{II}}$
because of the efficiency it has demonstrated in the description of
singlet states \citep{Piris2018a}, but the method proposed here can
be implemented to extent other approximate NOFs used in singlets.
It is important to remember that the pairing scheme of the orbitals
is allowed to vary along the optimization process until the most favorable
orbital interactions are found. Therefore, the orbitals do not remain
fixed in the optimization process, they adapt to the problem. In Fig.
\ref{fig1}, an illustrative example is shown. In this example, $S=1$
and $\mathrm{N_{I}}=2$, so two orbitals make up the subspace $\Omega_{\mathrm{I}}$,
whereas six electrons ($\mathrm{N_{II}}=6$) distributed in three
subspaces $\left\{ \Omega_{1},\Omega_{2},\Omega_{3}\right\} $ make
up the subspace $\Omega_{\mathrm{II}}$. The maximum value allowed
by the basis set is $\mathrm{N}_{g}=4$.

The reconstruction of the 2RDM must be subject to N-representability
conditions. In general, $D$ has a dependence on four indexes, computationally
expensive. We shall employ the two-index reconstruction proposed in
reference \citep{Piris2006}. This particular reconstruction is based
on the introduction of two auxiliary matrices $\Delta$ and $\Pi$
expressed in terms of the ONs to reconstruct the cumulant part of
the 2RDM. The necessary N-representability D, Q, and G conditions
of the 2RDM, also known as (2,2)-positivity conditions \citep{Mazziotti2012a},
impose strict inequalities on the off-diagonal elements of $\Delta$
and $\Pi$ matrices \citep{Piris2010a}. Appropriate forms of $\Delta$
and $\Pi$ have led to several $\mathcal{JKL}$-only functionals.
$\mathcal{J}$ and $\mathcal{K}$ refer to the usual Coulomb and exchange
integrals, $\left\langle pq|pq\right\rangle $ and $\left\langle pq|qp\right\rangle $
respectively, while $\mathcal{L}$ denotes the exchange-time-inversion
integral \citep{Piris1999} $\left\langle pp|qq\right\rangle $.

Let us extent the last member of this functional family, PNOF7 \citep{Piris2017,mitxelena2018a,Piris2018b},
to multiplets. Accordingly, divide the matrix elements of $D$ into
intra- and inter-subspace contributions. For intra-subspace blocks,
the functional form is maintained, that is, only intrapair $\alpha\beta$-contributions
appear,
\begin{equation}
\begin{array}{c}
D_{pq,rt}^{\alpha\beta}={\displaystyle \frac{\Pi_{pr}}{2}}\delta_{pq}\delta_{rt}\delta_{p\Omega_{g}}\delta_{r\Omega_{g}}\;(g\leq\frac{N_{\mathrm{II}}}{2})\quad\\
\\
\Pi_{pr}=\left\{ \begin{array}{c}
\sqrt{n_{p}n_{r}}\qquad p=r\textrm{ or }p,r>\frac{N_{\mathrm{II}}}{2}\\
-\sqrt{n_{p}n_{r}}\qquad p=g\textrm{ or }r=g\qquad\;
\end{array}\right.
\end{array}\label{intra}
\end{equation}
In (\ref{intra}), the Kronecker delta has an obvious meaning, for
instance, $\delta_{p\Omega_{g}}=1$ if $p\in\Omega_{g}$ or $\delta_{p\Omega_{g}}=0$
otherwise. Note also that $\Omega{}_{g}\in\Omega_{\mathrm{II}}$.
A major change is observed with respect to the $D^{\alpha\beta}$
of a singlet, namely: $D_{pp,pp}^{\alpha\beta}=0$, $\forall p\in\Omega_{\mathrm{I}}$.
Actually, there can be no interactions between electrons with opposite
spins in a singly occupied orbital, since there is only one electron
with $\alpha$ or $\beta$ spin in each $\left|SM\right\rangle $
of the ensemble.

A different ansatz is needed for inter-subspace contributions ($\Omega_{f}\neq\Omega{}_{g}$).
The spin-parallel matrix elements remain Hartree-Fock like,
\begin{equation}
D_{pq,rt}^{\sigma\sigma}={\displaystyle \frac{n_{p}n_{q}}{2}}\left(\delta_{pr}\delta_{qt}-\delta_{pt}\delta_{qr}\right)\delta_{p\Omega_{f}}\delta_{q\Omega_{g}},\:{}_{\sigma=\alpha,\beta}\label{interaa}
\end{equation}
whereas the spin-antiparallel blocks are
\begin{equation}
\begin{array}{c}
D_{pq,rt}^{\alpha\beta}={\displaystyle {\displaystyle \frac{n_{p}n_{q}}{2}}}\delta_{pr}\delta_{qt}\delta_{p\Omega_{f}}\delta_{q\Omega_{g}}-{\displaystyle \frac{\Phi_{p}\Phi_{r}}{2}}\delta_{p\Omega_{f}}\delta_{r\Omega_{g}}\cdot\\
\\
\cdot\left\{ \begin{array}{c}
\delta_{pq}\delta_{rt}\quad f\leq\frac{\mathrm{N_{II}}}{2}\textrm{ or }g\leq\frac{\mathrm{N_{II}}}{2}\\
\delta_{pt}\delta_{qr}\qquad\frac{\mathrm{N_{II}}}{2}<f,g\leq\mathrm{N}_{\Omega}\quad
\end{array}\right.
\end{array}\label{interab}
\end{equation}
where $\Phi_{p}=\sqrt{n_{p}(1-n_{p})}$. It should be noted that the
second term in Eq. (\ref{interab}) differs from our previous proposition
for singlets \citep{Piris2017,mitxelena2018a}. As a result, inter-subspace
$\alpha\beta$-contributions involving $\Phi$ in subspace $\Omega_{\mathrm{I}}$,
lead to exchange integrals instead of exchange-time-inversion integrals.

It is not difficult to verify that the reconstruction (\ref{intra})-(\ref{interab})
leads to $\mathrm{<}\hat{S}^{2}\mathrm{>}=\frac{\mathrm{N_{I}}}{2}\left(\frac{\mathrm{N_{I}}}{2}+1\right)$.
In fact, the expectation value of the operator $\hat{S}^{2}$ reads
as \citep{Piris2009}
\begin{equation}
\begin{array}{c}
\mathrm{<}\hat{S}^{2}\mathrm{>}={\displaystyle \frac{\mathrm{N}\left(4-\mathrm{N}\right)}{4}}{\displaystyle +\sum\limits _{pq}}\left\{ D_{pq,pq}^{\alpha\alpha}+D_{pq,pq}^{\beta\beta}-2D_{pq,qp}^{\alpha\beta}\right\} \end{array}\label{squareS}
\end{equation}
Note that
\[
\sum\limits _{pq}\quad\rightarrow\quad{\displaystyle \sum_{g=1}^{\mathrm{N}_{\Omega}}\;}{\displaystyle \sum_{p,q\in\Omega_{g}}+\sum_{f\neq g=1}^{\mathrm{N}_{\Omega}}\;\sum_{p\in\Omega_{f}}\sum_{q\in\Omega_{g}}}
\]

According to Eq. (\ref{interaa}), the spin-parallel matrix elements
of the 2RDM have only inter-subspace contributions ($\Omega_{f}\neq\Omega{}_{g}$).
Consequently, the trace of $D^{\sigma\sigma}$ can be cast as
\begin{equation}
\begin{array}{c}
\sum\limits _{pq}D_{pq,pq}^{\sigma\sigma}={\displaystyle \frac{1}{2}}{\displaystyle \sum_{f\neq g=1}^{\mathrm{N}_{\Omega}}\;\sum_{p\in\Omega_{f}}\sum_{q\in\Omega_{g}}}{\displaystyle n_{p}n_{q}}={\displaystyle \frac{1}{2}\left\{ {\displaystyle \sum_{f\neq g=1}^{\frac{\mathrm{N_{II}}}{2}}}+\right.}\\
\\
{\displaystyle \left.\sum_{f=1}^{\frac{\mathrm{N_{II}}}{2}}\;\sum_{g=\frac{\mathrm{N_{II}}}{2}+1}^{\mathrm{N}_{\Omega}}+\sum_{f=\frac{\mathrm{N_{II}}}{2}+1}^{\mathrm{\mathrm{N}_{\Omega}}}\;\sum_{g=1}^{\frac{\mathrm{N_{II}}}{2}}+\sum_{f\neq g=\frac{\mathrm{N_{II}}}{2}+1}^{\mathrm{N}_{\Omega}}\right\} }{\displaystyle \sum_{p\in\Omega_{f}}\sum_{q\in\Omega_{g}}}{\displaystyle n_{p}n_{q}}\\
\\
={\displaystyle \frac{1}{2}\left\{ {\displaystyle \sum_{f\neq g=1}^{\frac{\mathrm{N_{II}}}{2}}}1+\sum_{f=1}^{\frac{\mathrm{N_{II}}}{2}}\;\sum_{g=\frac{\mathrm{N_{II}}}{2}+1}^{\mathrm{N}_{\Omega}}\frac{1}{2}+\sum_{f=\frac{\mathrm{N_{II}}}{2}+1}^{\mathrm{\mathrm{N}_{\Omega}}}\;\sum_{g=1}^{\frac{\mathrm{N_{II}}}{2}}\frac{1}{2}\;+\right.}\\
\\
\left.{\displaystyle \sum_{f\neq g=\frac{\mathrm{N_{II}}}{2}+1}^{\mathrm{N}_{\Omega}}\frac{1}{4}}\right\} ={\displaystyle \frac{\mathrm{N_{II}\left(N_{II}-2\right)}}{8}+\frac{\mathrm{N_{II}N_{I}}}{4}+\frac{\mathrm{N_{I}\left(N_{I}-1\right)}}{8}}
\end{array}\label{Traa}
\end{equation}

where we have considered Eq. (\ref{sum1}), and 
\begin{equation}
{\displaystyle \sum_{p\in\Omega_{g}}}n_{p}=n_{g}=1/2,\,\Omega{}_{g}\in\Omega{}_{\mathrm{I}}
\end{equation}

Taking into account the $\alpha\beta$-blocks of the 2RDM in Eqs.
(\ref{intra}) and (\ref{interab}), the sum in the last $\alpha\beta$-term
of Eq. (\ref{squareS}) can be cast as

\begin{equation}
\begin{array}{c}
\sum\limits _{pq}D_{pq,qp}^{\alpha\beta}={\displaystyle \frac{1}{2}}{\displaystyle \sum_{g=1}^{\frac{\mathrm{N_{II}}}{2}}\;}{\displaystyle \sum_{p\in\Omega_{g}}}{\displaystyle n_{p}}-{\displaystyle \frac{1}{2}}{\displaystyle \sum_{f\neq g=\frac{\mathrm{N_{II}}}{2}+1}^{\mathrm{N}_{\Omega}}\;\sum_{p\in\Omega_{f}}\sum_{q\in\Omega_{g}}}{\displaystyle n_{p}n_{q}}\\
={\displaystyle \frac{1}{2}}\left\{ {\displaystyle {\displaystyle \sum_{g=1}^{\frac{\mathrm{N_{II}}}{2}}}\,1-\sum_{f\neq g=\frac{\mathrm{N_{II}}}{2}+1}^{\mathrm{N}_{\Omega}}\frac{1}{4}}\right\} ={\displaystyle \frac{\mathrm{N_{II}}}{4}}-{\displaystyle \frac{\mathrm{N_{I}\left(N_{I}-1\right)}}{8}}
\end{array}\label{abba}
\end{equation}

where it has been considered that $\Phi_{p}=\sqrt{n_{p}(1-n_{p})}=n_{p}=1/2$
for $\forall p\in\Omega_{\mathrm{I}}$.

Combining Eq. (\ref{squareS}) with Eqs. (\ref{Traa}) and (\ref{abba}),
one arrives at the ensemble average of the square of the total spin,
namely:
\begin{equation}
\begin{array}{c}
\mathrm{<}\hat{S}^{2}\mathrm{>}={\displaystyle \frac{\mathrm{\left(N_{I}+N_{II}\right)}\left(4-\mathrm{N_{I}-N_{II}}\right)}{4}}\\
\\
+2\left[{\displaystyle \frac{\mathrm{N_{II}\left(N_{II}-2\right)}}{8}+\frac{\mathrm{N_{II}N_{I}}}{4}+\frac{\mathrm{N_{I}\left(N_{I}-1\right)}}{8}}\right]\\
\\
{\displaystyle -\frac{\mathrm{N_{II}}}{2}}+{\displaystyle \frac{\mathrm{N_{I}\left(N_{I}-1\right)}}{4}}={\displaystyle \frac{\mathrm{N_{I}}}{2}\left(\frac{\mathrm{N_{I}}}{2}+1\right)}
\end{array}\label{S2}
\end{equation}

This result agrees with a multiplet of a total spin $S=\mathrm{N_{I}}/2$.
The corresponding multiplicity, $2S+1=\mathrm{N_{I}}+1$, is the highest
multiplicity for $\mathrm{N_{I}}$ electrons. A mixed state associated
to our approximate 2RDM reconstruction will be non-degenerate \citep{Paunez2000}.

In order to simplify the derivation, let us assume real spatial orbitals,
then $\mathcal{L}_{pq}=\mathcal{K}_{pq}$. After a simple algebra,
the energy (\ref{ENOF}) can be written as
\begin{equation}
E=\sum\limits _{g=1}^{\mathrm{N_{II}}/2}E_{g}+\sum\limits _{g=\mathrm{N_{II}}/2+\mathrm{1}}^{\mathrm{N}_{\Omega}}\mathcal{H}_{gg}+\sum\limits _{f\neq g}^{\mathrm{N}_{\Omega}}E_{fg}\label{EPNOF7}
\end{equation}
where
\begin{equation}
E_{g}=2\sum\limits _{p\in\Omega_{g}}n_{p}\mathcal{H}_{pp}+\sum\limits _{q,p\in\Omega_{g}}\Pi_{qp}\mathcal{K}_{pq}\,,\;\Omega{}_{g}\in\Omega_{\mathrm{II}}
\end{equation}
is the energy of a pair of electrons with opposite spins. It reduces
to a NOF obtained from ground-state singlet wavefunction, so $E_{g}$
describes accurately two-electron systems \citep{Piris2018a}. In
the last term of Eq. (\ref{EPNOF7}), $E_{fg}$ correlates the motion
of electrons with parallel and opposite spins belonging to different
subspaces ($\Omega_{f}\neq\Omega{}_{g}$):
\begin{equation}
E_{fg}=\sum\limits _{p\in\Omega_{f}}\sum\limits _{q\in\Omega_{g}}\left[n_{q}n_{p}\left(2\mathcal{J}_{pq}-\mathcal{K}_{pq}\right)-\Phi_{q}\Phi_{p}\mathcal{K}_{pq}\right]\label{Efg}
\end{equation}
The functional (\ref{EPNOF7})-(\ref{Efg}) is PNOF7 for multiplets.
\textcolor{black}{The solution is established by optimizing the energy
(}\ref{EPNOF7}\textcolor{black}{) with respect to the ONs and to
the NOs, separately. The conjugate gradient method is used for performing
the optimization of the energy with respect to auxiliary variables
that enforce automatically the N-representability bounds of the 1RDM.
The self-consistent procedure proposed in \citep{Piris2009a} yields
the NOs by an iterative diagonalization procedure, in which }orbitals
are not constrained to remain fixed along the orbital optimization
process.

\section{The NOF-MP2 method}

The weakness of approach (\ref{interab}) is the absence of the inter-subspace
dynamic electron correlation since $\Phi_{p}=\sqrt{n_{p}(1-n_{p})}$
has significant values only when the ONs differ substantially from
1 and 0. Consequently, PNOF7 is able to recover the complete intra-subspace,
but only the static inter-subspace correlation. To add the missing
dynamic correlation, we can use the NOF-MP2 method, Eq. (\ref{Etotal}).

The extension of NOF-MP2 to multiplets is straightforward and does
not require substantial changes with respect to the method implemented
for singlets \citep{Piris2018b}. PNOF7 provides the reference NOs
to form $\tilde{E}_{hf}$. However, one has to take into account the
presence of the singly occupied spatial orbitals. In this sense, the
zeroth-order Hamiltonian for the modified MP2 is constructed from
a closed-shell-like Fock operator that contains a HF density matrix
with doubly ($2n_{g}=2$) and singly ($2n_{g}=1$) occupied orbitals.
As a result, the reference energy $\tilde{E}_{hf}$ is
\begin{equation}
\tilde{E}_{hf}=2\sum\limits _{g=1}^{\mathrm{N}_{\Omega}}\mathcal{H}_{gg}+\sum_{f,g=1}^{\mathrm{N}_{\Omega}}\left(2\mathcal{J}_{fg}-\mathcal{K}_{fg}\right)-\sum_{g=\frac{\mathrm{N_{II}}}{2}+1}^{\mathrm{N}_{\Omega}}\frac{\mathcal{J}_{gg}}{4}\label{HFL}
\end{equation}
In Eq. (\ref{HFL}), the last term eliminates the $\alpha\beta$-contribution
of the singly occupied orbitals to the energy because in each pure
state $\left|SM\right\rangle $ of the ensemble there is no such interaction.

$E^{sta}$ is the sum of the static intra-space and inter-space correlation
energies:
\begin{equation}
\begin{array}{c}
E^{sta}=\sum\limits _{g=1}^{\mathrm{N_{II}/2}}\sum\limits _{q\neq p}\sqrt{\Lambda_{q}\Lambda_{p}}\,\Pi_{qp}\mathcal{\,K}_{pq}\\
\qquad-4\sum\limits _{f\neq g}^{\mathrm{N_{\Omega}}}\sum\limits _{p\in\Omega_{f}}\sum\limits _{q\in\Omega_{g}}\Phi_{q}^{2}\Phi_{p}^{2}\mathcal{K}_{pq}
\end{array}\label{Esta}
\end{equation}

where $\Lambda_{p}=1-\left|1-2n_{p}\right|$ is the amount of intra-space
static correlation in each orbital as a function of its occupancy.
Note that $\Lambda_{p}$ goes from zero for empty or fully occupied
orbitals to one if the orbital is half occupied.

$E^{dyn}$ is obtained from the second-order correction $E^{\left(2\right)}$
of the MP2 method. The first-order wavefunction is a linear combination
of all doubly excited configurations, considering one electron with
$\alpha$ or $\beta$ spin in $\Omega_{\mathrm{I}}$. The dynamic
energy correction takes the form
\begin{equation}
E^{dyn}=\sum\limits _{g,f=1}^{\mathrm{N_{\Omega}}}\;\sum\limits _{p,q>\mathrm{N}_{\Omega}}^{N_{B}}A_{g}A_{f}\left\langle gf\right|\left.pq\right\rangle \left[2T_{pq}^{gf}\right.\left.-T_{pq}^{fg}\right]\label{E2}
\end{equation}
where
\begin{equation}
A_{g}=\left\{ \begin{array}{c}
1,\quad1\leq g\leq\mathrm{N_{II}}/2\\
\dfrac{1}{2},\:\mathrm{N_{II}}/2<g\leq\mathrm{N}_{\Omega}
\end{array}\right.
\end{equation}

and $N_{B}$ is the number of basis functions. The amplitudes $T_{pq}^{fg}$
are obtained by solving the modified equations for the MP2 residuals
\citep{Piris2018b}.

In order to avoid double counting of the electron correlation, the
amount of dynamic correlation in each orbital $p$ is defined by functions
$C_{p}$ of its occupancy, namely,
\begin{equation}
\begin{array}{c}
C_{p}^{tra}=\begin{cases}
\begin{array}{c}
\begin{array}{c}
1-4\left(1-n_{p}\right)^{2}\end{array}\\
1-4n_{p}^{2}
\end{array} & \begin{array}{c}
p\leq\mathrm{N}_{\Omega}\\
p>\mathrm{N}_{\Omega}
\end{array}\end{cases}\\
\:C_{p}^{ter}=\begin{cases}
\begin{array}{c}
\begin{array}{c}
1\end{array}\\
1-4\left(1-n_{p}\right)n_{p}
\end{array} & \begin{array}{c}
p\leq\mathrm{N}_{\Omega}\\
p>\mathrm{N}_{\Omega}
\end{array}\end{cases}
\end{array}\label{Cp}
\end{equation}
where $C_{p}$ is divided into intra-space ($C_{p}^{tra}$) and inter-space
($C_{p}^{ter}$) contributions. According to Eq. (\ref{Cp}), fully
occupied and empty orbitals yield a maximal contribution to dynamic
correlation, whereas orbitals with half occupancies contribute nothing.
It is worth noting that $C_{p}^{ter}$ is not considered if the orbital
is below $\mathrm{N_{\Omega}}$. Using these functions as the case
may be (intra-space or inter-space), the modified off-diagonal elements
of the Fock matrix ($\tilde{\mathcal{F}}$) are defined as
\begin{equation}
\tilde{\mathcal{F}}_{pq}=\begin{cases}
C_{p}^{tra}C_{q}^{tra}\mathcal{F}_{pq}, & p,q\in\Omega_{g}\\
C_{p}^{ter}C_{q}^{ter}\mathcal{F}_{pq}, & otherwise
\end{cases}
\end{equation}
as well as modified two-electron integrals:
\begin{equation}
\widetilde{\left\langle pq\right|\left.rt\right\rangle }=\begin{cases}
C_{p}^{tra}C_{q}^{tra}C_{r}^{tra}C_{t}^{tra}\left\langle pq\right|\left.rt\right\rangle , & p,q,r,t\in\Omega_{g}\\
C_{p}^{ter}C_{q}^{ter}C_{r}^{ter}C_{t}^{ter}\left\langle pq\right|\left.rt\right\rangle , & otherwise
\end{cases}
\end{equation}
where the subspace index $g=1,...,\mathrm{N}_{\Omega}$. This leads
to the following linear equation for the modified MP2 residuals:
\begin{equation}
\widetilde{\left\langle ab\right|\left.ij\right\rangle }+\left(\mathcal{F}_{aa}\right.+\mathcal{F}_{bb}-\mathcal{F}_{ii}-\left.\mathcal{F}_{jj}\right)T_{ab}^{ij}\:+\label{residual}
\end{equation}
\[
{\displaystyle \sum_{c\neq a}\mathcal{\tilde{F}}_{ac}T_{cb}^{ij}}+{\displaystyle \sum_{c\neq b}}T_{ac}^{ij}\mathcal{\tilde{F}}_{cb}-{\displaystyle \sum_{k\neq i}}\tilde{\mathcal{F}}_{ik}T_{ab}^{kj}-{\displaystyle \sum_{k\neq j}}T_{ab}^{ik}\mathcal{\tilde{F}}_{kj}=0
\]
where $i,j,k$ refer to the strong occupied NOs, and $a,b,c$ to weak
occupied ones. It should be noted that diagonal elements of the Fock
matrix ($\mathcal{F}$) are not modified. By solving this linear system
of equations the amplitudes $T_{pq}^{fg}$ are obtained, which are
inserted into the Eq. (\ref{E2}) to achieve $E^{dyn}$.

\section{Ionization potentials of first-row transition metal atoms}

\begin{figure}
\centering{}\textcolor{black}{\includegraphics[scale=0.35]{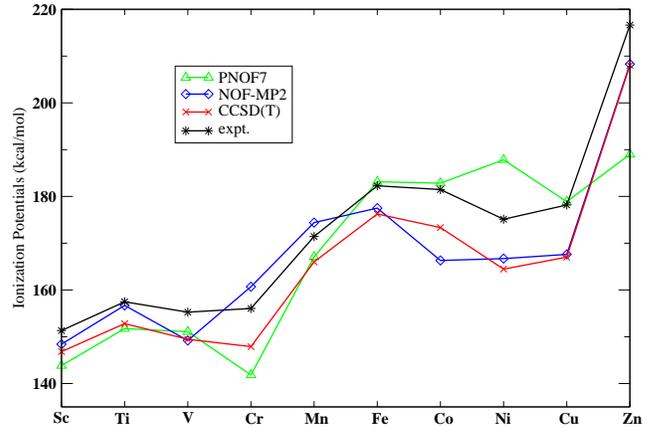}\caption{\label{fig2} Ionization potentials of the first-row transition-metal
atoms using cc-pVTZ basis set.}
}
\end{figure}

The ability to exist in multiple spin states makes transition-metal
(TM) atoms ideal candidates for testing PNOF7 in multiplets. Fig.
\ref{fig2} shows the calculated ionization potentials (IPs) of the
TM elements Sc-Zn using the correlation-consistent valence triple-basis
set including polarization (cc-pVTZ) \citep{Balabanov2005}. The IPs
are calculated by the energy difference between the positive ions
($\mathrm{X^{+}}$) and the neutral atoms (X): $\mathrm{IP}=E(\mathrm{X^{+}})-E(\mathrm{X})$.

For comparison, the IPs calculated using the same basis set at the
coupled cluster singles and doubles level of theory with perturbative
triples {[}CCSD(T)/cc-pVTZ{]}, and the experimental data reported
in Table IV of Ref. \citep{Balabanov2005}, have been included. The
data sets for these graphs can be found in Table \ref{tab:Comparison}.
It is worth noting that it is not intended to reproduce the experimental
data in this work, since it requires large basis sets and the inclusion
of relativistic effects.

\textcolor{black}{}

\begin{table}
\caption{\label{tab:Comparison}Ionizations potentials (kcal/mol) using the
cc-pVTZ basis set. CCSD(T) and experimental values taken from Table
IV of Ref. \citep{Balabanov2005}.\bigskip{}
\bigskip{}
}

\centering{}%
\begin{tabular}{clcccccccc}
\hline 
\multirow{1}{*}{Atom} &  & X &  & X$^{+}$ &  & PNOF7 & CCSD(T) & PNOF7-MP2 & EXP\tabularnewline
\hline 
Sc &  & $^{2}$D &  & $^{3}$D &  & 143.8 & 146.8 & 148.4 & 151.3\tabularnewline
Ti &  & $^{3}$F &  & $^{4}$F &  & 151.7 & 152.8 & 156.7 & 157.5\tabularnewline
V &  & $^{4}$F &  & $^{5}$D &  & 151.1 & 149.4 & 149.2 & 155.2\tabularnewline
Cr &  & $^{7}$S &  & $^{6}$S &  & 141.8 & 147.9 & 160.7 & 156.0\tabularnewline
Mn &  & $^{6}$S &  & $^{7}$S &  & 167.1 & 166.0 & 174.4 & 171.4\tabularnewline
Fe &  & $^{5}$D &  & $^{6}$D &  & 183.1 & 176.2 & 177.5 & 182.3\tabularnewline
Co &  & $^{4}$F &  & $^{3}$F &  & 182.8 & 173.3 & 165.7 & 181.5\tabularnewline
Ni &  & $^{3}$F &  & $^{2}$D &  & 187.8 & 164.5 & 166.7 & 175.1\tabularnewline
Cu &  & $^{2}$S &  & $^{1}$S &  & 178.9 & 167.0 & 167.6 & 178.2\tabularnewline
Zn &  & $^{1}$S &  & $^{2}$S &  & 189.0 & 208.0 & 208.3 & 216.6\tabularnewline
\hline 
MAE &  &  &  &  &  & 7.9 & 7.3 & 6.5 & \tabularnewline
\end{tabular}
\end{table}

There are small amounts of spin-contamination in the solution of the
CCSD equations \citep{Balabanov2005}. Nevertheless, let us take the
CCSD(T) values as benchmark data. The inspection of Fig. \ref{fig2}
reveals that calculated PNOF7 IPs from Sc to Mn are close to the CCSD(T)
values, but deviate from those for the late transition metals, reaching
deviations of 23 and 19 kcal/mol for Ni and Zn, respectively. Note
that the mean absolute error (MAE) with respect to the experiment
is quite similar for both methods, 7.9 and 7.6 kcal/mol respectively,
although the behavior of PNOF7 is somewhat erratic due to the aforementioned
systems.

The results obtained for the IPs of the TM elements Sc-Zn at the PNOF7-MP2
level of theory have also been included in Fig. \ref{fig2}. It is
important to note that in the case of the neutral atoms Co, Ni, and
Cu, as well as in the Co and Ni ions, there are two solutions. One
of these solutions is the most favorable energetically at the PNOF7
level, whereas the other solution corresponds to the lowest in energy
for the NOF-MP2 method. In the former, the non-dynamic character of
the electron correlation predominates, while the dynamic correlation
predominates in the latter. This reflects the complexity of the problem
in these systems where the 3d shell is almost filled.

The data reveal a remarkable improvement, especially in the Cr, Ni
and Zn atoms. The inclusion of inter-subspace dynamic electron correlation
yields a MAE of 6.5 kcal/mol, which is an outstanding result considering
the size of the basis sets employed. Large basis sets could improve
the MAE to values lower than 1 kcal/mol \citep{Balabanov2005}.

\textbf{\textcolor{black}{Acknowledgments:}}\textcolor{black}{{} Financial
support comes from MCIU/AEI/FEDER, UE (PGC2018-097529-B-100). The
author thanks for technical and human support provided by IZO-SGI
SGIker of UPV/EHU and European funding (ERDF and ESF).}

\end{document}